\documentclass[twocolumn,aps,noshowpacs,preprintnumbers,amsfonts,amsmath,amssymb,superscriptaddress,floatfix]
{revtex4}
\usepackage[dvipdfm]{graphicx}
\usepackage{amsmath}	% required for `\align' (yatex added)

\newcommand{\ket}[1]{\left\vert #1 \right\rangle}

\newcommand{\reffig}[1]{Fig.~\ref{#1}}
\newcommand{\expect}[1]{\langle #1 \rangle}
\renewcommand{\a}{\hat{a}}
\newcommand{\vac}{\ket{0}}
\newcommand{\e}{\mathrm{e}}
\newcommand{\x}{\hat{x}}
\newcommand{\p}{\hat{p}}
\newcommand{\half}{\frac{1}{2}}

\let \dag=\dagger
\let \lam=\lambda
\let \mr=\mathrm
\let \mc=\mathcal
\newcommand{\affA}{%
\affiliation{
     National Institute of Information and Communications Technology
     (NICT), \\
     4-2-1 Nukui-kitamachi, Koganei, Tokyo 184-8795, Japan}
     }
\newcommand{\affB}{%
\affiliation{
	Department of Applied Physics and Quantum Phase Electronics Center,
	School of Engineering,
	The University of Tokyo,\\
	7-3-1 Hongo Bunkyo-ku, Tokyo 113-8656, Japan}
	}

\begin{document}
%
%\title{Experimental entanglement distillation from Gaussian states}
%\title{Beyond the no-go bound for the Gaussian entanglement distillation\\ or: How I learned to stop worrying and love the bound}
\title{Non-Gaussian entanglement distillation for continuous variables}
\date{\today}

\author{Hiroki Takahashi}
\affA
\affB

\author{Jonas S. Neergaard-Nielsen}
\author{Makoto Takeuchi}
\author{Masahiro Takeoka}
\author{Kazuhiro Hayasaka}
\affA
\author{Akira Furusawa}
\affB
\author{Masahide Sasaki}%
\affA%
%

%\pacs{42.50.Dv, 42.50.Ex, 03.67.-a, 03.65.Wj}

%42.50.Dv Quantum state engineering and measurements
%42.50.Ex Optical implementations of quantum information processing
%and transfer
%03.67.-a Quantum information
%03.65.Wj State reconstruction, quantum tomography

\begin{abstract}
Entanglement distillation is an essential ingredient %in the quantum repeater
for long distance quantum communications
\cite{briegel98:_quant_repeat}.
%It also has been of fundamental interest in quantum information theory \cite{horodecki07:_quant_entan}.
% For discrete variable systems entanglement distillation using photonic
% \cite{kwiat01:_exper_entan_distil_and_hidden_non_local,
% pan01:_entan_purif_for_quant_commun,
% yamamoto03:_exper_extrac_of_entan_photon} and atomic
% \cite{reichle06:_exper_purif_of_two_atom_entan} qubits have been
% demonstrated.
%On the contrary
In the continuous variable setting, Gaussian states play major roles
%formally analogous to qubit states
in quantum teleportation, quantum cloning and quantum cryptography
\cite{braunstein05:_quant_infor_with_contin_variab}. However,
entanglement distillation from Gaussian states has not yet been
demonstrated. It is made difficult by the no-go theorem stating that no
Gaussian operation can distill Gaussian states
\cite{eisert02:_distil_gauss_states_with_gauss,
fiuraifmmode02:_gauss_trans_and_distil_of,
giedke02:_charac_of_gauss_operat_and}. Here we demonstrate the
entanglement distillation from Gaussian states by using
measurement-induced non-Gaussian operations, circumventing the
fundamental restriction of the no-go theorem. We observed a gain of
entanglement as a result of conditional local subtraction of a single
photon or two photons from a two-mode Gaussian state.
%At the maximal gain the logarithmic negativity was more than doubled.
Furthermore  we confirmed that two-photon subtraction also improves Gaussian-like
entanglement as specified by the Einstein-Podolsky-Rosen (EPR)
correlation. This distilled entanglement can be further employed to
downstream applications such as high fidelity quantum teleportation
\cite{opatrny00:_improv_telep_of_contin_variab} and a loophole-free
Bell test \cite{garcia-patron04:_propos_for_looph_free_bell}. 
\end{abstract}

\maketitle

Long distance quantum communications rely on the ability to faithfully
distribute entanglement between distant locations.
%possibly by photons through fiber networks or via a satellite.
However, inevitable decoherence and the inability to amplify quantum
signals hinder efforts to extend a quantum optical link to the
practically large scale.
To overcome this difficulty, %qunatum repeater was proposed. There,
entanglement distillation can be used %between neighboring nodes plays a fundamental role.
- a protocol in which each distant party locally manipulates particles
of less entangled pairs with the aid of classical communication to
extract a smaller number of pairs of higher entanglement
\cite{bennett96:_purif_of_noisy_entan_and}. In discrete variable
systems many distillation experiments have already been demonstrated
\cite{kwiat01:_exper_entan_distil_and_hidden_non_local,pan01:_entan_purif_for_quant_commun,yamamoto03:_exper_extrac_of_entan_photon,reichle06:_exper_purif_of_two_atom_entan}.

An alternative to the discrete variable system is the one described by
continuous variables (CV), typically represented by
%the position and momentum of a harmonic oscillator and
the quadrature amplitudes of a light field. For CV quantum information,
Gaussian states and Gaussian operations
\cite{ferraro05:_gauss_states_in_contin_variab_quant_infor} are of
particular importance. They are readily available in the laboratory and
serve as a complete framework for many quantum protocols
\cite{braunstein05:_quant_infor_with_contin_variab}.
The first experiments of CV entanglement distillation were reported in
\cite{hage08:_prepar_of_distil_and_purif,dong08:_exper_entan_distil_of_mesos_quant_states}.
These two works rely on Gaussian operations. However, it was
theoretically proven that Gaussian operations can never distill
entanglement from Gaussian state inputs -- this is known as the no-go
theorem of Gaussian operations
\cite{eisert02:_distil_gauss_states_with_gauss,fiuraifmmode02:_gauss_trans_and_distil_of,giedke02:_charac_of_gauss_operat_and}.
In
\cite{hage08:_prepar_of_distil_and_purif,dong08:_exper_entan_distil_of_mesos_quant_states},
the inputs had been subjected to some specific classes of non-Gaussian
noise, such as phase-diffusion
\cite{hage08:_prepar_of_distil_and_purif} or temporally varying
attenuation \cite{dong08:_exper_entan_distil_of_mesos_quant_states}. In
those cases, well established Gaussian technologies can be applied to
distill the entanglement.

So far, there has been no demonstration of entanglement distillation with Gaussian inputs.
This task essentially requires a new technology of non-Gaussian operations.
Recent theories also revealed that this is a must to realize quantum
speed-up of CV quantum information processing (QIP)
\cite{bartlett02:_effic_class_simul_of_contin}. Triggered by this new
paradigm of non-Gaussian QIP, the research field extending to the
non-Gaussian regime has rapidly developed
\cite{ourjoumtsev06:_gener_optic_schroed_kitten_for,neergaard-nielsen06:_gener_of_super_of_odd,wakui07:_photon_subtr_squeez_states_gener}.
The increase \cite{ourjoumtsev07:_increas_entan_between_gauss_states}
and preparation \cite{2009NatPh...5..189O} of entanglement from
Gaussian inputs by {\it non-local} photon subtraction were also
demonstrated. These are important steps towards the realization of
entanglement distillation from Gaussian states.

\begin{figure}[tb]
 \includegraphics[width=\linewidth]{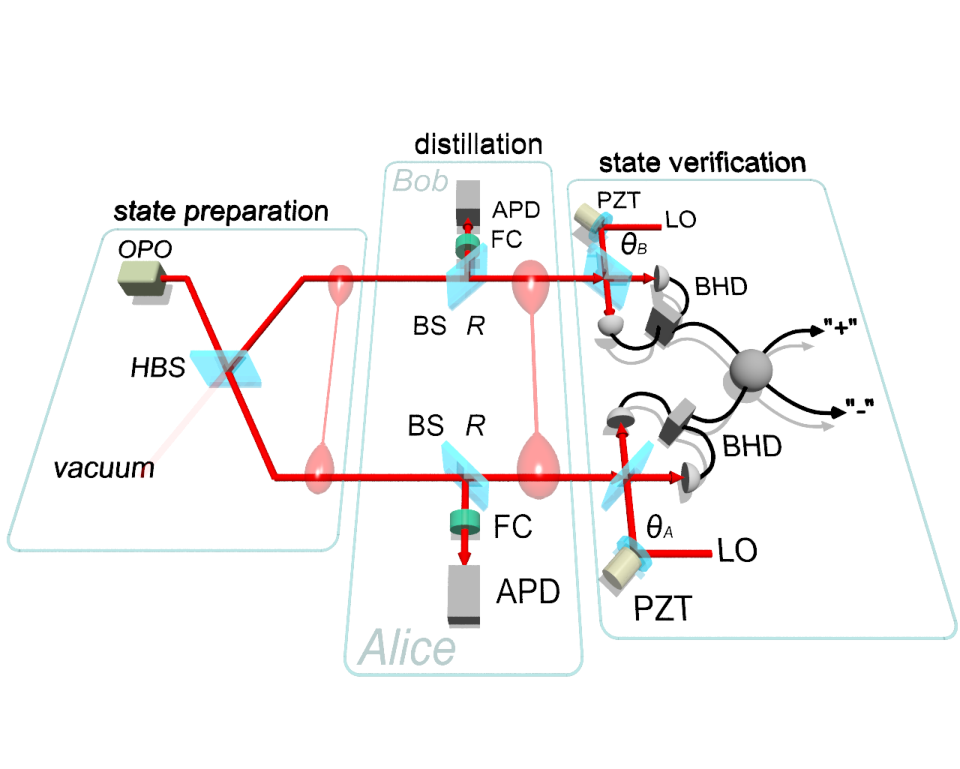}
 \caption{
 Schematic of the experiment. OPO: optical parametric oscillator,
 %HWP: half wave plate, PBS: polarization beam splitter,
 BS: beam splitter, HBS: half-beam splitter,
 BHD: balanced homodyne detector,
 LO: local oscillator, FC: filtering cavity,
 APD: avalanche photo diode, PZT: Piezo electric transducer.
 %Beam splitters except for the ones  in homodyne detection are realized by HWP and PBS.
 The phases of the LOs, $\theta_A$ and  $\theta_B$ are stabilized by electronic feedback
 to the PZTs through field programmable gate array (FPGA) modules.
 The outputs of the BHDs are recorded by a digital oscilloscope triggered
 by either logical ``AND'' or ``OR'' of the click signals from the two
 APDs.
 For the state verification, a set of the homodyne outcomes are
 numerically converted into the ``$+/-$'' basis.
 }
 \label{fig:schematic}
\end{figure}

Here we report on the entanglement distillation directly from CV
Gaussian states by using local photon subtraction as non-Gaussian
operations, circumventing the no-go restriction on Gaussian operations.
A schematic of our experiment is depicted in \reffig{fig:schematic}.
A continuous wave squeezed vacuum is generated from an optical parametric oscillator
(OPO) detailed elsewhere \cite{wakui07:_photon_subtr_squeez_states_gener}.
The initial Gaussian entangled state is prepared by splitting the squeezed vacuum by half at
the first beam splitter and is distributed to the separate parties, Alice and Bob.
This half-split squeezed vacuum with squeezing parameter $r$ is effectively equivalent to the two-mode squeezed vacuum
with $r/2$ (they are compatible by local unitary operations. see Appendix \ref{sec:local-unit-equiv}).
%, and thus is regarded as a decohered entangled pair ({\bf ???}).
% A local, non-Gaussian operation -- photon subtraction --
% is performed probabilistically by either one or both of the parties.
At each site of Alice and Bob, a probabilistic non-Gaussian operation
-- photon subtraction -- is performed. Specifically, a small part of
the beam is picked off by a polarizing beam splitter with the variable
reflectance $R$ and sent through filtering cavities
\cite{wakui07:_photon_subtr_squeez_states_gener} to an avalanche
photodiode (APD) to detect a photon (\reffig{fig:schematic}). Each
photon detection at the APD heralds a local success of the photon
subtraction attempt. Conditioned on the subtraction of a photon by a
single party (single-photon subtraction) or the simultaneous
subtraction of a photon by both parties (two-photon subtraction), Alice
and Bob retain those two-mode states which have successfully had their
entanglement increased. While the single-photon subtraction scheme will
have a higher success rate, the two-photon subtraction scheme will give
a more Gaussian-like final state which is more readily applicable to
further processing such as e.g. quantum teleportation.

The distillation works since the local photon subtraction changes the
non-local unfactorizable correlations of the initial state. To see this
intuitively, let us describe the initial squeezed state with squeezing parameter $r$ in photon number basis
as $\ket{0,0} - \frac{\lambda}{2\sqrt{2}} (\ket{0,2} + \sqrt{2}\ket{1,1} + \ket{2,0}) +
O(\lambda^2)$ where $|m,n\rangle = |m\rangle_A |n\rangle_B$, $\lambda =
\tanh r$ and we omitted the normalization. For small $\lambda$, it is
almost factorizable since $\ket{0,0}$ is dominant.
%In other words, a flatter distribution of the superpostions provides
%higher entanglement.
Applying the single-photon subtraction, represented by an annihilation
operator $\hat{a}$, the state is transformed to be $\sqrt{2}(\ket{0,1}
+ \ket{1, 0}) + O(\lambda)$ which is clearly more entangled - the first
Bell state term corresponds to 1 ebit of entanglement. We emphasize
that the term $O(\lambda)$ is not negligible for larger $\lambda$ and
critically contributes to the distillation, and thus the scheme works
for any $\lambda$ in principle. See Appendix \ref{sec:entangl-dist-stat} for the
rigorous formulation and for the two-photon subtraction.

The verification of the distilled (or undistilled) states is carried
out by a quantum tomographic method with two local homodyne
measurements (see \reffig{fig:schematic}). For tomography, in general the
local oscillator (LO) phases $\theta_A$ and $\theta_B$ (for Alice and
Bob's detectors, respectively) have to be swept over all possible
combinations to collect full information of the two-mode state. In our
case, however, it is significantly simplified due to the fact that our
states, distilled or undistilled, are always represented in a
decomposed way as
\begin{align}
 \label{eq:wig-fact}
 W(x_A, p_A, x_B, p_B)& = W_s(x_-, p_-)W_v(x_+, p_+),
\end{align}
where $W$ is the Wigner function for the two-mode state, $x_A, x_B,
p_A, p_B$ are the quadrature amplitudes of modes $A$ and $B$,
$x_{\pm}=\frac{x_A\pm x_B}{\sqrt{2}}$, $p_{\pm}=\frac{p_A\pm
p_B}{\sqrt{2}}$ and $W_s$, $W_v$ are the Wigner functions for a (zero-,
one-, or two-) photon subtracted squeezed vacuum and the vacuum
respectively. For such a state, the scans of the homodyne measurements
are necessary only for $\theta_A = \theta_B$ and the experimental data
of $x_{\pm}$ is numerically obtainable from the measured $x_A$ and
$x_B$ (\reffig{fig:schematic}).
%This property holds even when the initial squeezed vacuum would become a mixed state and $R$ is finite (see Method).
It should be stressed that although we assume the state factorization of (\ref{eq:wig-fact}),
it can be directly assessed by verifying experimentally whether the state of the g+h mode is indeed
a pure vacuum state. For details, see Appendix \ref{sec:fact-dist-stat}.
%Note that a similar situation was studied in \cite{ourjoumtsev07:_increas_entan_between_gauss_states}. but importantly in our case the assumption of factorization can be directly assessed by verifying whether the state of the``$+$'' mode is indeed the pure vacuum state.

\begin{figure*}[tb]
 \includegraphics[width=1\linewidth]{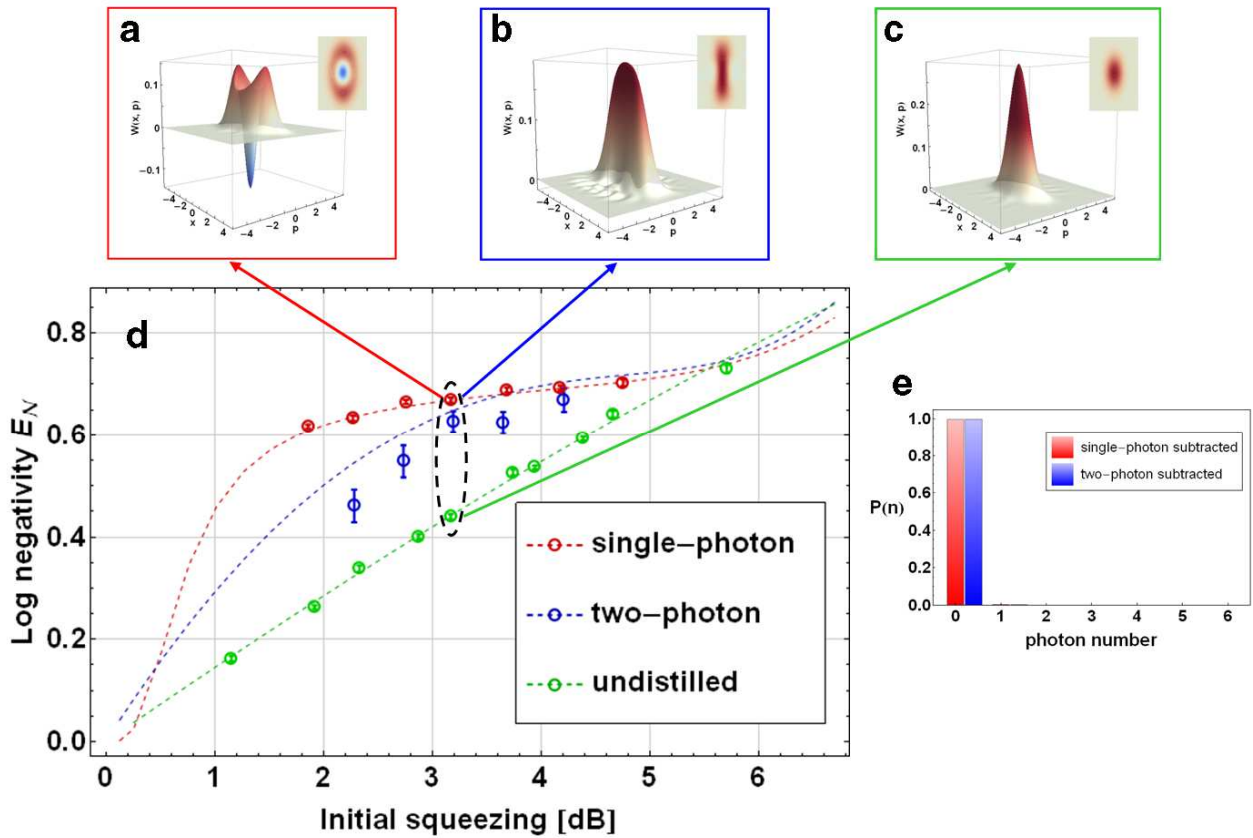}
 \caption{
 \textbf{a, b, c}
 Experimentally reconstructed Wigner functions and their contour plots
 of the ``$-$'' mode states for (a) Distilled state via single-photon
 subtraction with $R=$5\%, (b) Distilled state via two-photon
 subtraction with $R=$10\%, and (c) Undistilled initial state (squeezed
 vacuum with $R=$0\%), all with initial squeezing of -3.2 dB.
 \textbf{d} Experimental logarithmic negativities as functions of the initial input
 squeezing. Here the initial squeezing refers to the
 squeezing of the states right after the OPO and is deduced from
 separately measured classical parametric amplification of the OPO. For
 the single-photon subtracted states (red) and the undistilled states
 (green), 600,000 samples over 6 phases were used for the
 reconstruction of each point. For the two-photon subtracted states
 (blue) 18,000 - 48,000 samples were used. The dashed curves are
 theoretical predictions based on independently  measured experimental
 parameters.
 Every error bar represents an uncertainty of the state reconstruction and was estimated via a MonteCarlo simulation using the corresponding experimental parameters.
 \textbf{e}, The photon number distributions of the  experimentally reconstructed ``$+$'' mode states corresponding to (a)
 and (b). }
 \label{fig:wig}
\end{figure*}

% We carried out the complete homodyne tomography of the single-photon subtracted two-mode state corresponding to the one with ($n_A$, $n_B$)=(1, 0) in
% (\ref{eq:2mode-state-vec}).
% Marginal quadrature distributions for 36 different combinations of phases at Alice and Bob's site were recorded.
% Both phases were stabilized by using a time-binned auxiliary coherent beam as in \cite{takahashi08:_gener_of_large_amplit_coher} but an efficient computer-based automation was
% additionally implemented with custom FPGA modules.
Examples of reconstructed Wigner functions obtained by the single- and two-photon subtractions, as well as
the initial squeezed state are shown in \reffig{fig:wig}a-c.
The outputs of the homodyne detectors were
sampled at 6 different phases of LO, namely $\theta_A = \theta_B =0,
\pi/6, \pi/3, \pi/2, 2\pi/3, 5\pi/6$. We extract the measured values of
the quadratures $\hat{x}_A$ and $\hat{x}_B$ by applying a mode function
to the recorded traces
\cite{neergaard-nielsen06:_gener_of_super_of_odd,
wakui07:_photon_subtr_squeez_states_gener,
takahashi08:_gener_of_large_amplit_coher}.
After calculating the corresponding values of
$\hat{x}_\pm$, we reconstruct the density matrices for the ``$+$'' and
``$-$'' modes  by the conventional maximum likelihood estimation
\cite{lvovsky04:_iterat_maxim_likel_recon_in} without any correction of
detection losses.

As shown in \reffig{fig:wig}e, for the ``$+$'' mode states we got
almost perfectly pure vacuum states with more than 99\% accuracy. We
confirmed that this holds irrespective of the initial squeezing level.
This experimental evidence justifies our tomography scheme based on the
relation (\ref{eq:wig-fact}). On the other hand, for the ``$-$'' mode
states we observed two different kinds of non-Gaussian state depending
on whether single photon or two photons were subtracted
(\reffig{fig:wig}a and b). They respectively correspond to the
odd and even Schr\"{o}dinger cat state, i.e.
$\ket{\alpha}-\ket{-\alpha}$ and $\ket{\alpha}+\ket{-\alpha}$ where
$\ket{\alpha}$ is a coherent state with coherent amplitude $\alpha$.
Having these reconstructed states we can use the relation
(\ref{eq:wig-fact}) backwards to calculate the amount of entanglement
shared by Alice and Bob. Specifically, we calculate the logarithmic
negativity $E_\mathcal{N}$ which is a monotone measure of entanglement
\cite{vidal02:_comput_measur_of_entan}.
\begin{figure}[tb]
 \includegraphics[width=\linewidth]{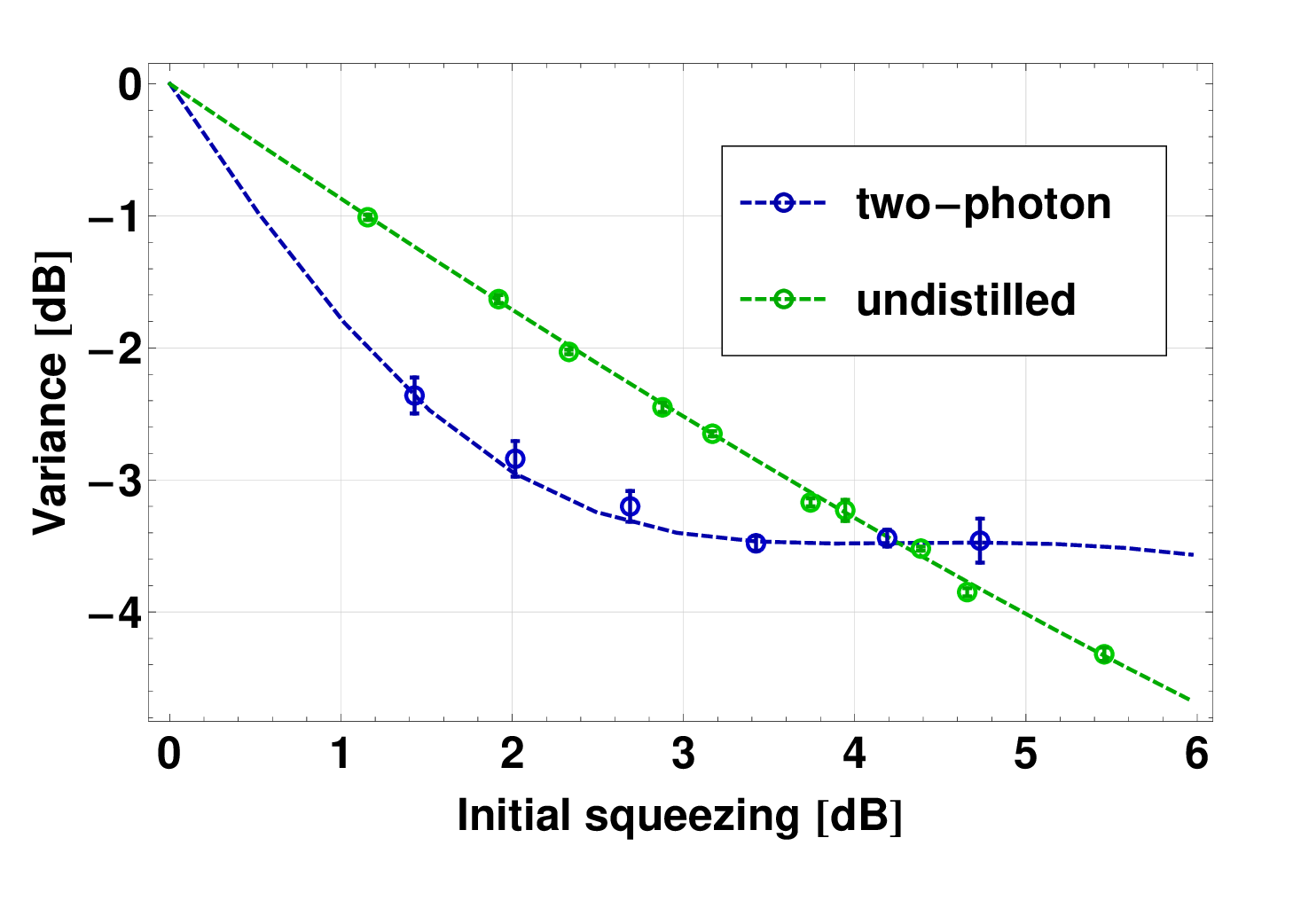}
 \caption{ Squeezed variances of $x_-$ (normalized by the vacuum
 level). For the undistilled states each point was calculated from the
 full-reconstructed density matrices common to the negativity
 measurements. For the distilled states each point was obtained by
 directly measuring variance of 1,600 - 5,000 samples at the most
 squeezed phase.
%by using 1,600 - 5,000 samples.
 } \label{fig:neg-EPR}
\end{figure}

\reffig{fig:wig}d  shows the experimental negativities of the
undistilled Gaussian states, the states distilled by single-photon
subtraction with $R=5\%$, and by two-photon subtraction with $R=10\%$
as functions of the squeezing of the initial input states. When
evaluating negativity, one must take care of its strong dependency on
the size of the data set. We investigated the behavior of the
negativity on the data size and deduced an extrapolative value
corresponding to an infinitely large data set for each point in
\reffig{fig:wig}d (See Appendix \ref{sec:calc-logar-negat}). Note that without this analysis,
evaluation of negativity with finite sized data very likely goes into
an overestimate of the negativity. As shown in the figure, over a wide
range of the initial squeezing we got clear gains of entanglement
relative to the undistilled Gaussian states for both the single- and
two-photon subtracted schemes.

A practical difference between the single- and
two-photon subtracted scheme is on their rates of event detection. In
the single-photon experiment the rate is around a few thousands per
second, but in the two-photon experiment there are only a few events per
second. So while for the former we can use hundreds of thousands of
samples for the state reconstruction, for the latter we can only use
a few tens of thousands limited by the long-term stability of the setup.
In \reffig{fig:wig}d the experimental negativities for the single-photon subtracted states and the undistilled states are in
very good agreement with theory, but ones for the two-photon subtraction are slightly below the theoretical predictions.
This may be due to an uncontrollable drift of the system during a long period of the measurements.

As can be seen in \reffig{fig:wig}d, in terms of the logarithmic
negativity the two-photon subtracted scheme does not have an advantage
over the single-photon subtracted scheme despite its significantly
lower success rate. However the two-photon subtracted distillation
transforms a two-mode Gaussian state into one relatively close to a
Gaussian state (see \reffig{fig:wig}b). Hence one would expect that
states distilled by this scheme still possess a Gaussian-like property
of entanglement. For Gaussian states, two-mode entanglement is usually
specified in terms of the Einstein-Podolsky-Rosen (EPR) correlation
quantified by $\expect{(\Delta \hat{x}_-)^2}\expect{(\Delta\hat{p}_+)^2}$.
Since the ``$+$'' mode is always a vacuum state (see Eq.
\ref{eq:wig-fact}), we can focus on the degree of squeezing of the
``$-$'' mode as an equivalent measure. We carried out measurements of
the variance of $\hat{x}_-$ at its most squeezed phase conditioned on
two-photon subtraction (~\reffig{fig:neg-EPR}). Note that this
measurement is considerably faster than reconstruction of a full state
and is possibly more accurate due to its simplicity. The results in
\reffig{fig:neg-EPR} show that the two-photon subtraction improves the
EPR correlation of a state with up to 4 dB of initial squeezing. An
improvement on the EPR correlation gives us an operational measure of
success of the two-photon subtracted distillation in terms of the
fidelity of CV quantum teleportation
\cite{opatrny00:_improv_telep_of_contin_variab}. On the other hand, the
single-photon subtracted distillation is never able to improve the
degree of the EPR correlation. This fact demonstrates a clearly
different nature of the distilled entanglement between the two schemes.

In conclusion, we have for the first time demonstrated CV entanglement
distillation from Gaussian states by conditional local photon
subtraction. Because of the importance of Gaussian states and the no-go
theorem of the Gaussian distillation,
% in CV quantum information,
this has been a long-standing experimental milestone to be achieved.
Our scheme would serve as the de-Gaussifying process of a more generic
distillation protocol proposed in
\cite{eisert04:_distil_of_contin_variab_entan} by combining it with the
already demonstrated Gaussification processes
\cite{hage08:_prepar_of_distil_and_purif,dong08:_exper_entan_distil_of_mesos_quant_states},
which would realize long-distance CV quantum communications. Finally,
our non-Gaussian entangled states are not only useful for
communications but also for fundamental problems such as a loophole-free test of Bell's inequality
\cite{garcia-patron04:_propos_for_looph_free_bell}.

H.T. acknowledges the financial support from G-COE program commissioned
by the MEXT of Japan.

%Our scheme would be the first de-Gaussifying process of the iterative
%distillation protocol in
%\cite{eisert04:_distil_of_contin_variab_entan}.
%
%Also our highly non-Gaussian entangled state would be used for a
%loophole-free Bell test
%\cite{garcia-patron04:_propos_for_looph_free_bell} or quantum
%teleportation.
%
%It can also serve as a starting point of a more generic CV entanglement
%distiller by combining it with the already demonstrated Gaussification
%protocols
%\cite{hage08:_prepar_of_distil_and_purif,dong08:_exper_entan_distil_of_mesos_quant_states}.
%We observed clear evidences of distilled entanglement in terms of the logarithmic negativity in the cases of the single- and two-photon subtraction.
%In the case of the two-photon subtraction we also observed improvements on the EPR correlation between two modes.
% \begin{figure}[tb]
%  \includegraphics[width=\linewidth]{figs/EPR.eps}
% % \caption{}
%  \label{fig:epr}
% \end{figure}

\appendix

\section{Theoretical description of the distilled states}

\subsection{Factorizability of the distilled states}
\label{sec:fact-dist-stat}

\begin{figure*}[tbh]
 \includegraphics[width=.8\linewidth]{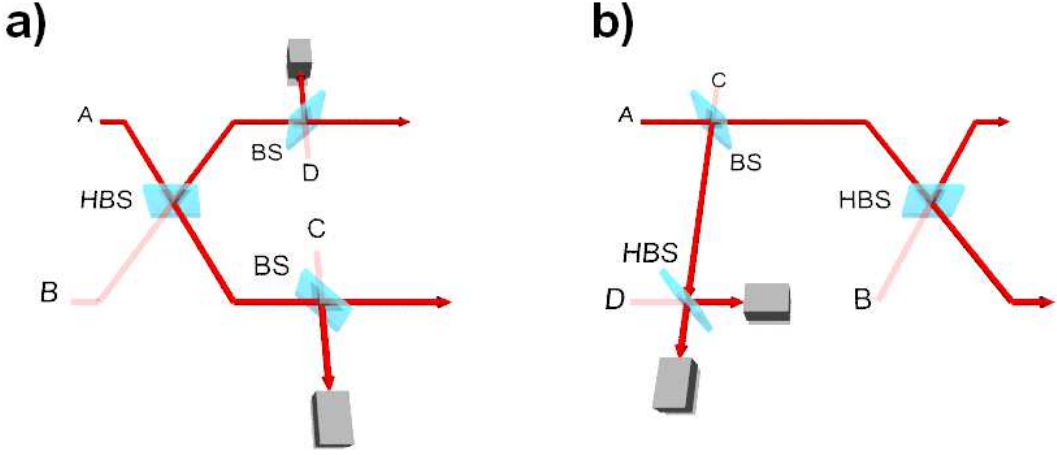}
 \caption{
 {\bf a} Schematic of the local photon subtraction.
 BS: beam splitter, HBS: half beam splitter. The letters A, B, C and D label the optical modes.
 The thinner lines represent the vacuum inputs. The gray boxes are photon detectors. Here the state of mode A is half-split and then photons are subtracted locally from the outputs of the HBS.
 {\bf b} An equivalent model for a. The state of mode A is firstly photon-subtracted and then half-split with mode B.
 }
 \label{fig:model}
\end{figure*}

Here we discuss the local photon subtraction from a two-mode entangled Gaussian state made by splitting a
squeezed vacuum.
\reffig{fig:model}a illustrates this scheme,
where one of the inputs for the half-beam splitter is in the vacuum state.
Conditioning on either photon detection at one of the detectors or coincidence detection of two detectors
heralds a non-Gaussian two-mode state effectively described as
follows (apart from the normalization).
\begin{align}
 \ket{\psi_{out}}_{AB}&=\a_A^{n_A}\a_B^{n_B}\hat{B}_{AB}(\pi/4) \hat{S}_A(r)\vac_A\vac_B \label{eq:ideal-photon-sub} \\
 &=\hat{B}_{AB}(\pi/4)\,\a_A^{n_A+n_B}\hat{S}_A(r)\vac_A\vac_B, \label{eq:2mode-state-vec}
\end{align}
where $\hat{S}_A(r)$ is the squeezing operator with the squeezing
parameter $r$ and $\hat{B}_{AB}(\pi/4)$ is the half-beam splitter
operator for the modes $A$ and $B$. More generally we denote the beam
splitter operator with reflectance $R$ as $\hat{B}_{AB}(\theta)$ where
$\theta=\sin^{-1}(\sqrt{R})$ and
$\hat{B}_{AB}(\theta)=\exp(\theta(\a^\dag_A\a_B-\a_A\a^\dag_B)$). The
integers $n_A$ and $n_B$ represent the number of photons subtracted,
i.e.  ($n_A$, $n_B$)=(1, 0) or (0, 1) for the single photon detection
at either site and ($n_A$, $n_B$)=(1, 1) for the coincidence detection
of both sites. The state given by (\ref{eq:2mode-state-vec}) is a
half-split state of the photon subtracted squeezed vacuum. Hence the
situation described by \reffig{fig:model}a can be equivalently
transformed to the one in \reffig{fig:model}b. Although the description
of (\ref{eq:ideal-photon-sub}) for photon subtraction is only valid in
the limit where the reflectance of the local tapping BS goes to zero,
as will be shown in the following, the equivalency between
\reffig{fig:model}a and b still holds even when the photon subtraction
is not ideal, i.e. the reflectance of the BS is not infinitesimal.

Consider the situation in \reffig{fig:model}a again, with definitions
of modes A, B, C, and D. Here we assume that the reflectance of the
tapping BS is finite. The following operator identity holds:
\begin{align}
 &\hat{B}_{AC}(\theta)\hat{B}_{BD}(\theta)\hat{B}_{AB}(\pi/4) \nonumber\\
 &=\hat{B}_{AB}(\pi/4)\hat{B}_{AB}^\dag(\pi/4)\hat{B}_{AC}(\theta)\hat{B}_{BD}(\theta)\hat{B}_{AB}(\pi/4) \\
 &=\hat{B}_{AB}(\pi/4)\hat{B}_{AB}^\dag(\pi/4) \nonumber\\
 &\qquad\times\exp\left(\theta(\a^\dag_A\a_C-\a_A\a^\dag_C+\a^\dag_B\a_D-\a_B\a^\dag_D)\right)
 \hat{B}_{AB}(\pi/4) \\
 &=\hat{B}_{AB}(\pi/4)
 \exp\left(\theta
 \left(-\a_A\frac{\a^\dag_C-\a^\dag_D}{\sqrt{2}}+\a^\dag_A\frac{\a_C-\a_D}{\sqrt{2}} \right.\right. \nonumber \\
 &\qquad\left.\left.-\a_B\frac{\a^\dag_C+\a^\dag_D}{\sqrt{2}}+\a^\dag_B\frac{\a_C+\a_D}{\sqrt{2}}
 \right)
 \right) \\
 &=\hat{B}_{AB}(\pi/4)\hat{B}_{CD}(\pi/4)\hat{B}_{AC}(\theta)\hat{B}_{BD}(\theta)\hat{B}_{CD}^{\dag}(\pi/4),
\end{align}
where $\hat{B}_{ij}$ stands for the beam splitter operator for modes
$i$ and $j$. Then for an arbitrary input state $\ket{\Psi}$ for mode A
and the vacuum states for modes B, C, and D:
\begin{align}
 &\hat{B}_{BD}(\theta)\hat{B}_{AC}(\theta)\hat{B}_{AB}(\pi/4)\ket{\Psi}_A\vac_B\vac_C\vac_D
 \nonumber\\
 &=\hat{B}_{AB}(\pi/4)\hat{B}_{CD}(\pi/4)\hat{B}_{AC}(\theta)\hat{B}_{BD}(\theta)\hat{B}_{CD}^{\dag}(\pi/4) \nonumber \\
 &\qquad\times\ket{\Psi}_A\vac_B\vac_C\vac_D \\
 &=\hat{B}_{AB}(\pi/4)\hat{B}_{CD}(\pi/4)\hat{B}_{AC}(\theta) \ket{\Psi}_A\vac_B\vac_C\vac_D
\end{align}
This establishes the equivalency between the models shown in
\reffig{fig:model}a and b again. Therefore the final output state of
the protocol is identical to a half-split of the photon subtracted
squeezed vacuum.

Let us return to the form of (\ref{eq:2mode-state-vec}) for simplicity.
Then we immediately have
\begin{align}
 \hat{B}^\dag(\pi/4)\ket{\psi_{out}}_{AB}&=\a_A^{n_A+n_B}\hat{S}_A(r)\vac_A\vac_B,
 \label{eq:dist-disentangle}
\end{align}
which means that if we let the conditional two-mode state be combined
at a half-beam splitter it becomes disentangled and furthermore the
outputs get separated as a photon subtracted squeezed vacuum and
the vacuum. Introducing new variables
\begin{align}
 \x_\pm=\frac{\x_A\pm\x_B}{\sqrt{2}} \quad \p_\pm=\frac{\p_A\pm\p_B}{\sqrt{2}},
 \label{eq:dist-def-x-p-pm}
\end{align}
where $\x_{A, B}$ and $\p_{A, B}$ are the quadrature observables for Alice and Bob's subsystem, then from the identity (\ref{eq:dist-disentangle}),
in terms of these variables the Wigner function of the output state can be written as
\begin{align}
 W_{out}(x_A, p_A, x_B, p_B)&=W_{v}(x_+, p_+)W_{s}(x_-, p_-) \label{eq:dist-disentangle-wig},
\end{align}
where $W_{v}$ and $W_{s}$ are the Wigner functions of the vacuum state
and the photon subtracted squeezed vacuum state respectively.

% The remaining two-mode state after conditioning is analyzed by two
% local homodyne measurements (see \reffig{fig:setup}). In general, by
% sweeping the local oscillator (LO) phases $\theta_A$ and $\theta_B$
% over their all possible combinations while measuring quadrature
% amplitudes, we can collect enough information for the two-mode state
% reconstruction. However for our specific state of
% (\ref{eq:2mode-state-vec}) the tomographic measurement can be further
% simplified. As seen from the right hand side of
% (\ref{eq:2mode-state-vec}), the state generated is equivalent to a
% half-split of a photon subtracted squeezed vacuum.
% %if we recombine the two beams of Alice and Bob by the operator $\hat{B}_{AB}^{\dag}$ the state would become separable with the
% %single-mode photon subtracted squeezed vacuum, a close approximation of a Schr\"{o}dinger kitten state,  at one hand and the vacuum at the other.
% %In other words,
% Therefore the Wigner function of the two-mode state can be written as %a form of

\subsection{Local unitary equivalency between a half-split squeezed vacuum and a two-mode squeezed vacuum}
\label{sec:local-unit-equiv}

A two-mode Gaussian state with zero local displacements is completely specified by its covariance matrix given by
\begin{widetext}
\begin{equation}
 \begin{pmatrix}
 \expect{\x_A^2} & \frac{1}{2}{\expect{\x_A\p_A+\p_A\x_A}} & \frac{1}{2}{\expect{\x_A\x_B+\x_B\x_A}} & \frac{1}{2}{\expect{\x_A\p_B+\p_B\x_A}} \\
 \frac{1}{2}{\expect{\x_A\p_A+\p_A\x_A}} & \expect{\p_A^2} & \frac{1}{2}{\expect{\p_A\x_B+\x_B\p_A}} & \frac{1}{2}{\expect{\p_A\p_B+\p_B\p_A}} \\
 \frac{1}{2}{\expect{\x_A\x_B+\x_B\x_A}} & \frac{1}{2}{\expect{\p_A\x_B+\x_B\p_A}} & \expect{\x_B^2} & \frac{1}{2}{\expect{\x_B\p_B+\p_B\x_B}} \\
 \frac{1}{2}{\expect{\x_A\p_B+\p_B\x_A}} & \frac{1}{2}{\expect{\p_A\p_B+\p_B\p_A}} & \frac{1}{2}{\expect{\p_A\p_B+\p_B\p_A}}  & \expect{\p_B^2}
\end{pmatrix}.
\end{equation}
\end{widetext}
The covariance matrix of the half-split squeezed vacuum $\ket{\Psi_0} = \hat{B}_{AB}(\pi/4)\hat{S}_A(r)\vac_A\vac_B$ is
 \begin{align}
 \begin{pmatrix}
 \frac{\e^{-r}\cosh(r)}{2} & 0 & \frac{\e^{-r}\sinh(r)}{2}& 0 \\
 0 & \frac{\e^{r}\cosh(r)}{2} &  0 & \frac{-\e^{r}\sinh(r)}{2} \\
 \frac{\e^{-r}\sinh(r)}{2} & 0 & \frac{\e^{-r}\cosh(r)}{2}& 0 \\
 0 & \frac{-\e^{r}\sinh(r)}{2} &  0 & \frac{\e^{r}\cosh(r)}{2}
 \end{pmatrix}. \nonumber
 \end{align}
Performing local squeezing operations on both modes, it can be made to
have symmetric variances:
 \begin{align}
 \begin{pmatrix}
 \frac{\cosh(r)}{2} & 0 & \frac{\sinh(r)}{2}& 0 \\
 0 & \frac{\cosh(r)}{2} &  0 & \frac{-\sinh(r)}{2} \\
 \frac{\sinh(r)}{2} & 0 & \frac{\cosh(r)}{2}& 0 \\
 0 & \frac{-\sinh(r)}{2} &  0 & \frac{\cosh(r)}{2}
 \end{pmatrix}. \nonumber
 \end{align}
This is identical to the covariance matrix of a two-mode squeezed state
with squeezing parameter $\frac{r}{2}$. So in terms of entanglement,
the half-split squeezed vacuum with squeezing parameter $r$ is
equivalent to the two-mode squeezed state with squeezing parameter
$\frac{r}{2}$. By using a state vector, this is described as
 \begin{align}
 \hat{S}_A(-r/2)\hat{S}_B(-r/2)\ket{\Psi_0}
 =(1-\lam'^2)^\half\sum_{n=0}^\infty \lam'^{\,n}\ket{n}_A\ket{n}_B, \label{eq:equiv-hssv-2msv}
 \end{align}
with $\lam'=\tanh{r/2}$.

\subsection{Entanglement of the distilled states}
\label{sec:entangl-dist-stat}

Entropy of entanglement is defined as the von Neumann entropy of a
reduced subsystem. Every bipartite pure state can be brought into the
form of the Schmidt decomposition with Schmidt coefficients $c_n$:
\begin{align}
 \sum_{n=0}^{\infty}\sqrt{c_n}\ket{n}_A\ket{n}_B.
\end{align}
From this, the entropy of the subsystem is calculated as
\begin{align}
 E &=-\mr{tr}(\rho_A\log(\rho_A)) = -\mr{tr}(\rho_B\log(\rho_B) \\
 &= -\sum_n c_n\log{c_n}. \label{eq:entropy-formula}
\end{align}
From (\ref{eq:equiv-hssv-2msv}) and the fact that local operations do not alter the amount of entanglement, we get the Schmidt coefficients of the half-split squeezed
vacuum $\ket{\Psi_0}$ as
 \begin{align}
  c_n^{(0)} & = (1-\lam'^2) \lam'^{2n}. \label{eq:hssv-schmidt}
 \end{align}

\begin{figure}[thb]
 \includegraphics[width=\linewidth]{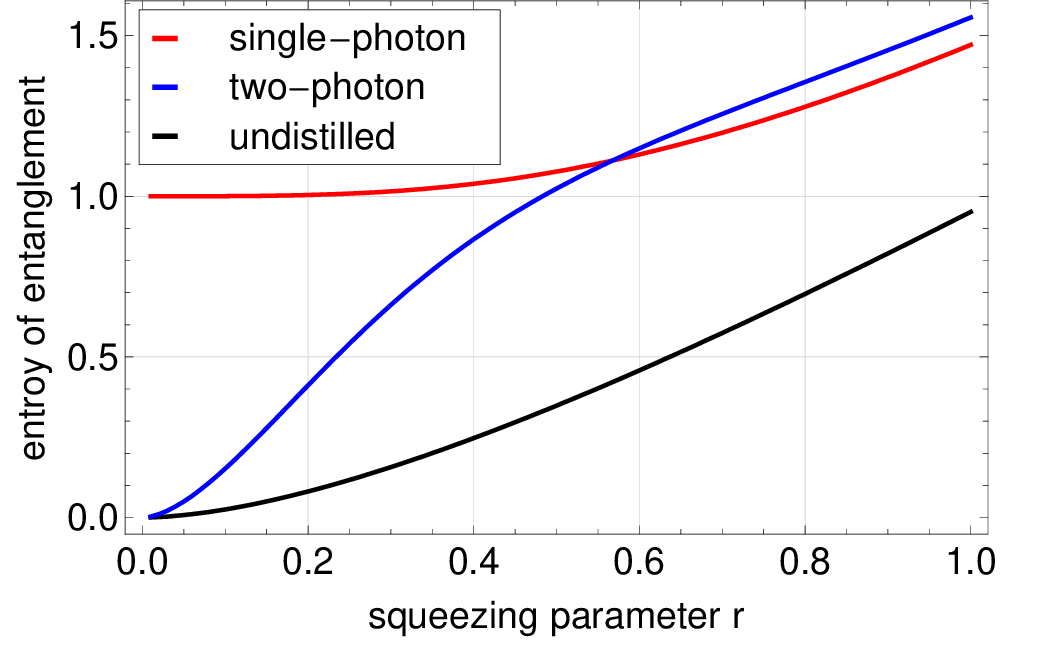}
 \caption{The entropy of entanglement for the single-photon subtracted states (red), the two-photon subtracted states (blue) and the undistilled states (black) as functions
 of the squeezing parameters $r$.
 }
  \label{fig:vNent}
\end{figure}

Let us denote the single-photon subtracted half-split squeezed vacuum as $\ket{\Psi_1}$.   
\begin{align}
 \ket{\Psi_1} &= \frac{1}{\sqrt{\mc{N}_1}}\a_A\ket{\Psi_0}, \\
 \mc{N}_1 &=\frac{\sinh^2{r}}{2}.
\end{align}
From the argument in the last section, we have
 \begin{align}
 &\hat{S}_A(-r/2)\hat{S}_B(-r/2)\ket{\Psi_1} \nonumber \\
 &=\frac{1}{\sqrt{\mc{N}_1}}\hat{S}_A(-r/2)\hat{S}_B(-r/2)\a_A\ket{\Psi_0} \\
 &=\frac{1}{\sqrt{\mc{N}_1}}(\a_A\cosh(r/2)+\a^\dag_A\sinh(r/2)) \nonumber \\
 &\qquad\qquad\times\hat{S}_A(-r/2)\hat{S}_B(-r/2)\ket{\Psi_0} \\
 &=\frac{1}{\sqrt{\mc{N}_1}}(\a_A\cosh(r/2)+\a^\dag_A\sinh(r/2))(1-\lam'^2)^\half \nonumber\\
 &\qquad\qquad\times\sum_{n=0}^\infty \lam'^{\,n}\ket{n}_A\ket{n}_B \\
 &=\frac{(1-\lam'^2)^\half}{\sqrt{\mc{N}_1}} \sum_{n=0}^\infty \lam'^{\,n}
 (\cosh(r/2)\sqrt{n}\ket{n-1}_A \nonumber \\
 &\qquad\qquad+\sinh(r/2)\sqrt{n+1}\ket{n+1}_A)\ket{n}_B \\
 &=\frac{(1-\lam'^2)^\half}{\sqrt{\mc{N}_1}} \sum_{m=0}^\infty \sum_{n=0}^\infty A_{mn}\ket{m}_A\ket{n}_B,
 \end{align}
where
 \begin{align}
 A_{mn} &= \lam'^{\,n}(\cosh(r/2)\sqrt{n}\delta_{m, n-1} \nonumber \\
  &\qquad+\sinh(r/2)\sqrt{n+1}\delta_{m, n+1}). 
 \end{align}
From the singular values of matrix $A_{mn}$ (let them be $\alpha_n$), the Schmidt coefficients for $\ket{\Psi_1}$ are given by
 \begin{equation}
 c^{(1)}_n=\frac{(1-\lam'^2)}{\mc{N}_1}\alpha_n^2. \label{eq:single-photon-schimidt}
 \end{equation}

Similar to the single photon case, we obtain a representation of the
two-photon subtracted half-split squeezed vacuum $\ket{\Psi_2}$ in the
following form.
 \begin{align}
 &\hat{S}_A(-r/2)\hat{S}_B(-r/2)\ket{\Psi_2} \nonumber \\
 &= \frac{1}{\sqrt{N}_2}\hat{S}_A(-r/2)\hat{S}_B(-r/2)\a_A\a_B\ket{\Psi_0} \\
 &= \frac{(1-\lam'^2)}{\sqrt{N}_2}
 \sum_{m=0}^\infty \sum_{n=0}^\infty B_{mn}\ket{m}_A\ket{n}_B,
 \end{align}
where
 \begin{align}
 \mc{N}_2 &=\frac{2\sinh^4{r}+\cosh^2{r}\sinh^2{r}}{4} \\
 B_{mn} &= \cosh^2(r/2)(m+1)\lam'^{\,m+1}\delta_{m, n} \nonumber \\
 &+\sinh^2(r/2) m\lam'^{\,m-1}\delta_{m, n} \nonumber \\
 &+\cosh(r/2)\sinh(r/2)\sqrt{(m+1)(m+2)} \lam'^{\,m+1}\delta_{m+2, n} \nonumber \\
 &+\cosh(r/2)\sinh(r/2)\sqrt{m(m-1)} \lam'^{\,m-1}\delta_{m-2, n}
 \end{align}
Then the Schmidt coefficients $c^{(2)}_n$ for $\ket{\Psi_2}$ are  given by the singular values $\beta_n$ of matrix $B_{mn}$:
\begin{align}
 c^{(2)}_n & = \frac{(1-\lam'^2)}{\mc{N}_2}\beta_n^2 \label{eq:two-photon-schmidt}
\end{align}
(\ref{eq:hssv-schmidt}), (\ref{eq:single-photon-schimidt}), (\ref{eq:two-photon-schmidt}) and (\ref{eq:entropy-formula}) we can readily calculate the entropy of
entanglement for those states. 
\reffig{fig:vNent} shows the entropy of entanglement for the single-photon subtracted states $\ket{\Psi_1}$, the two-photon subtracted states $\ket{\Psi_2}$
 and the undistilled states $\ket{\Psi_0}$ as functions  of squeezing parameter $r$.
Note that in an actual experiment, we have several experimental imperfections and we end up with a mixed state output.
In such cases, the pure state descriptions above no longer  hold and the entropy of entanglement is not a good measure of entanglement,
but \reffig{fig:vNent} still outlines the general behavior of this protocol. 

\section{Calculating the logarithmic negativity}
\label{sec:calc-logar-negat}

In principle, the logarithmic negativity $E_\mathcal{N}$ can be
directly calculated when we know the 2-mode entangled state, as the sum
of the negative eigenvalues of the partially transposed density matrix,
$(\rho_{AB})^{T_A}$ \cite{vidal02:_comput_measur_of_entan}. In the
experiment, however, we found that $E_\mathcal{N}$ is very sensitive to
statistical noise in the measurements. Smaller data sets lead to larger
errors in the reconstructed density matrix elements which ultimately
leads to larger calculated $E_\mathcal{N}$ values. The intuitive
understanding of this effect is the following: The absolute errors on
each density matrix element due to statistical measurement noise are
roughly the same. Hence, the relative errors are large for the
high-photon number elements which are all close to zero - most likely
their absolute values will increase due to the errors. But high photon
numbers contribute a significant amount to the overall entanglement of
the state, so in the end more noise will give seemingly higher
entanglement.

\begin{figure}[ht]
 \includegraphics[width=\linewidth]{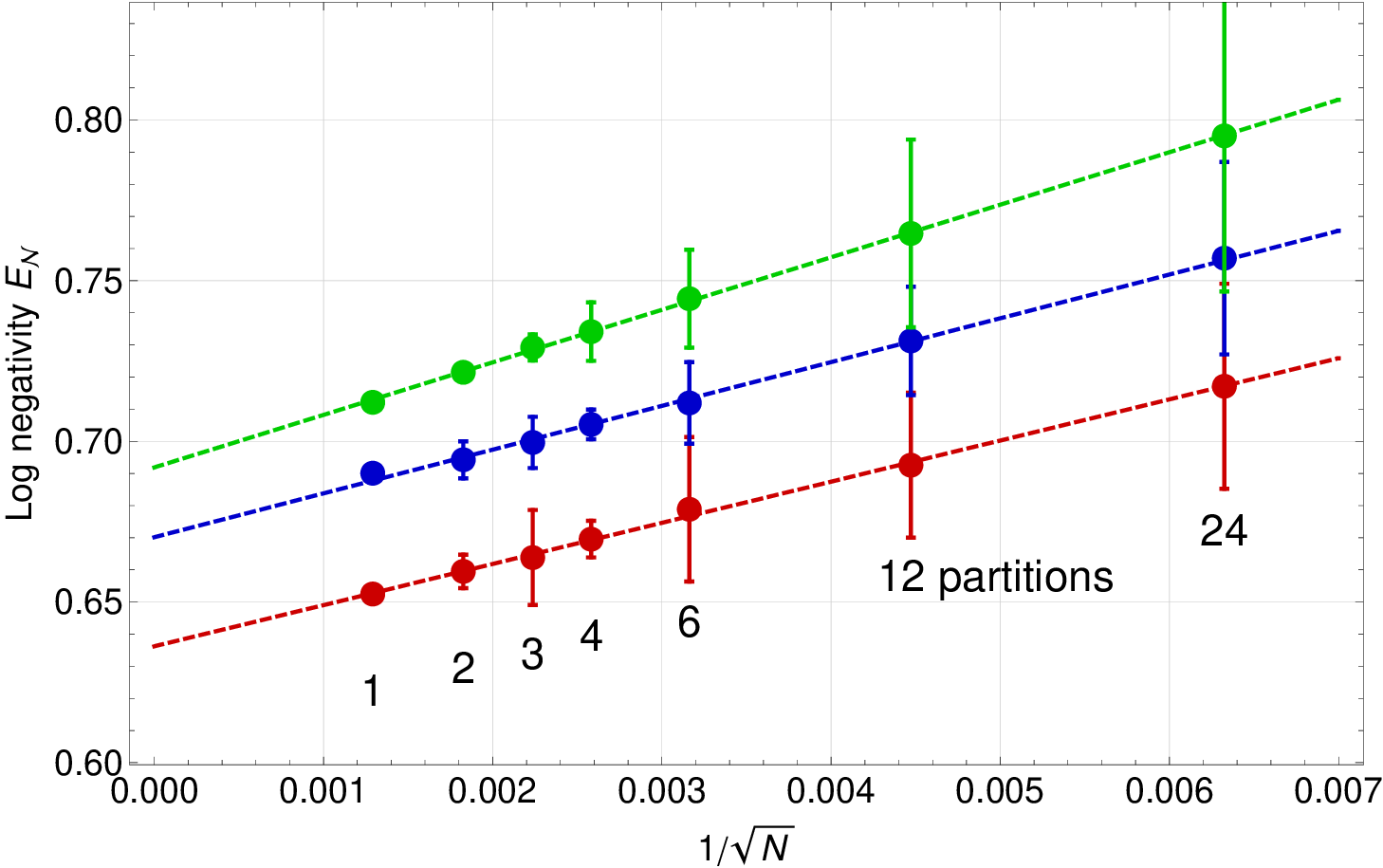}
 \caption{Estimation of the logarithmic negativity for infinitely many
 data samples.
 %The green (upper) points come from experimental data,
 %while the blue (lower) points are based on simulated data where we
 %know the true value of $E_{\mathcal{N}}$ (red).
 Each point shows the average and standard deviation of the calculated
 logarithmic negativities, $E_\mathcal{N}(N_d)$ (average taken over the
 $d$ subsets of the full data set), versus $1/\sqrt{N_d}$, where $N_d$
 is the size of each subset. The points are well fitted by a line whose
 y-axis intersection (infinite data size) gives a good estimate for the
 true $E_{\mathcal{N}}$ value (as confirmed by simulated data). The
 three plot series are from single-photon subtracted data sets with
 different initial squeezing levels, 2.3 (red), 3.2 (blue), and 4.2 (green) dB respectively.
 %The total size of both the experimental and
 %simulated data were 600,000 samples. For simulated data of this size,
 %the estimated $E_{\mathcal{N}}(\infty)$ value is usually within 2\%
 %deviation from the true value. Note that the simulation presented here
 %is not supposed to emulate the measured data -- the chosen initial
 %squeezing level was different from the -4.7 dB of the measured state.
 }
  \label{fig:asympnegplot}
\end{figure}

We found empirically that the calculated negativity scales with the
total data sample size $N$ as $E_\mathcal{N}(N) = a + b/\sqrt{N}$. We
interpret this as $E_\mathcal{N}(\infty)=a$ being the ``true'' value
that we would obtain in the asymptotic limit of very large data sample
size, while the second term is the contribution from statistical noise.
To obtain this value $E_\mathcal{N}(\infty)$ from a given data set of
$N_{\mathrm{full}}$ samples, %$X_{full}=\{(x_{A,B})_i\}, i=1,\ldots,N_{full}$,
we perform multiple state reconstructions based on truncations of the
full data set. Specifically, we partition %$X_{full}$
the full set into $d$ subsets %$X_j^{(d)}$
of $N_d\simeq N_{\mathrm{full}}/d$ samples each (with equal
representation of all phase angles). The logarithmic negativity is then
calculated from the reconstruction of all $d$ subsets, and we take the
mean value of these to be an estimate for $E_{\mathcal{N}}(N_d)$. We
repeat the process for other numbers of partitions, $d$, and thereby
get a plot as in Fig. \ref{fig:asympnegplot} of the dependency of
calculated entanglement on data sample size. A least-squares fit to $a
+ b/\sqrt{N}$ then gives us the asymptotic estimate
$E_\mathcal{N}(\infty)$. We have confirmed by simulated data that this
approach does in fact give the correct value for $E_{\mathcal{N}}$
within roughly $\pm 2\%$.

\bibliography{ref}

\end{document}